\documentclass[reprint,twocolumn,aps,pra,amsmath,amssymb,floatfix,superscriptaddress,longbibliography]{revtex4-1}

\usepackage{graphicx}
\usepackage{epsfig}
\usepackage{bm}
\usepackage{dcolumn}
\usepackage{color}
\usepackage[colorlinks,urlcolor=cyan,citecolor=blue,linkcolor=magenta]{hyperref}
\makeatletter
\newcommand*{\rom}[1]{\expandafter\@slowromancap\romannumeral #1@}
\makeatother
\usepackage{tikz}
\usetikzlibrary{calc,decorations, decorations.markings, decorations.pathreplacing, decorations.pathmorphing, decorations.shapes, decorations.text}

\begin{document}

\title{Nonlinear Hall response in the driving dynamics of ultracold atoms in optical lattices}
\author{Xiao-Long Chen}
\affiliation{Department of Physics, Zhejiang Sci-Tech University, Hangzhou 310018, China}
\affiliation{Institute for Advanced Study, Tsinghua University, Beijing 100084, China}
\author{Wei Zheng}
\email{zw8796@ustc.edu.cn}
\affiliation{Hefei National Laboratory for Physical Sciences at the Microscale and Department of Modern Physics, University of Science and Technology of China,
Hefei 230026, China}
\affiliation{CAS Center for Excellence in Quantum Information and Quantum Physics, University of Science and Technology of China, Hefei 230026, China}
\date{\today }

\begin{abstract}
We propose that a nonlinear Hall response can be observed in Bloch oscillations of ultracold atoms in optical lattices under the condition of preserved time-reversal symmetry. In the short-time limit of Bloch oscillations driven by a direct current (dc) field, the nonlinear Hall current dominates, being a second-order response to the external field strength. The associated Berry curvature dipole, which is a second-order nonlinear coefficient of the driving field, can be obtained from the oscillation of atoms. In an alternating current (ac) driving field, the nonlinear Hall response has a double frequency of the driving force in the case of time-reversal symmetry.
\end{abstract}

\maketitle

\section{Introduction}
The Hall effect plays an important role in condensed matter physics~\cite{hall1879new}. It is commonly used to measure the charge of carriers in conductors. Its quantized version, the quantum Hall effect, was observed in two-dimensional electron gases in 1980~\cite{klitzing1980new}. The quantization of Hall conductance is determined by the fundamental topological properties of materials, such as the Berry curvature and gapless boundary states, which are robust against disorder and impurity content~\cite{xiao2010berry,nagaosa2010anomalous,cage2012quantum}. Over the past few decades, there have been intensive studies that broadened the quantum Hall family, from the fractional quantum Hall effect~\cite{tsui1982two}, the quantum anomalous Hall effect~\cite{chang2013experimental}, to topological insulators~\cite{hasan2010colloquium,qi2011topological}. These studies reshaped our understanding of the phases and phase transitions of materials.

Time-reversal symmetry (TRS) is broken in both the Hall effect and anomalous Hall effect, by an external magnetic field and by spontaneous symmetry breaking respectively. The linear-order transverse response to an applied electric field, i.e. the Hall conductance, vanishes in the presence of TRS. However, systems with TRS can exhibit a nonvanishing Hall response beyond the linear order, e.g., to the quadratic-order external electric field. This nonlinear Hall effect (NLHE) was proposed by Sodemann and Fu in materials with broken inversion symmetry (IS)~\cite{sodemann2015quantum}. Later, it was intensively studied in condensed matter systems~\cite{deyo2009semiclassical,moore2010confinement,low2015topological,facio2018strongly,you2018berry,zhang2018berry,du2018band,du2019disorder,du2021perspective}. Very recently this effect was observed in multilayered WTe$_{2}$ structures~\cite{ma2019observation,kang2019nonlinear} and topological insulators of Bi$_{2}$Se$_{3}$~\cite{he2019nonlinear}.
These studies revealed that NLHE can be induced by intrinsic and extrinsic factors. The extrinsic factors are caused by disorder~\cite{du2019disorder}, while the intrinsic one is related to the Berry curvature and relies on the nonequilibrium distribution of carriers in the Bloch band. In condensed matter systems, when the backscattering from impurities balances the action of the applied electric field, the carriers reach a current carrying steady state. The distribution of carriers is slightly shifted from the equilibrium one. Under these conditions, the intrinsic NLHE can be detected. In systems with cold atoms, the external electrical field is replaced by a gradient potential, in view of the electric neutrality of carriers, i.e., atoms. Due to the lack of backscattering from impurities, atoms will exhibit continuous, undamped Bloch oscillations under the action of an external force~\cite{dahan1996bloch,anderson1998macroscopic}. The distribution of carriers is far from the equilibrium one  during the oscillation and evolves in time. It is natural to ask if the NLHE can be observed in such far-from-equilibrium Bloch oscillations.

In this paper, we study the nonlinear Hall response during Bloch oscillations of ultracold atoms in an optical lattice. Under the action of a dc driving force, the dynamics of the transverse current is dominated by the second-order nonlinear Hall effect in a short time in systems with TRS but without IS. The Berry curvature dipole, which reflects the distribution of Berry curvature in the Brillouin zone, can be obtained from the transverse velocity of atoms. It is worth mentioning that, the Berry curvature in hexagonal optical lattices can be now reconstructed directly by measuring the band wave functions in cold-atom experiments~\cite{flaschner2016experimental,li2016bloch}. In the case of an ac driving force, the transverse drifts will oscillate with a half period of the driving force in a time-reversal symmetric band. This is a typical frequency doubling induced by the nonlinear Hall effect. On the other hand, the period of transverse oscillations in systems with broken TRS is identical to the one of the driving force. The Berry curvature dipole can be also obtained from the amplitude of the Hall response.

The rest of this paper is organized as follows. The Bloch oscillation and semiclassical approximation are introduced in Sec.~\ref{sec:blochoscillations}. In the presence of an external dc field, we first study the short-time dynamical behaviors of the mean velocity (see Fig.~\ref{fig2}) and of the Berry curvature dipole (see Fig.~\ref{fig3}). We then consider the long-time dynamics in both dc and ac cases by studying the mean velocity and measurable displacement (see Figs.~\ref{fig4}-\ref{fig6}). A summary and outlook are given in Sec.~\ref{sec:summary}.

\section{Bloch oscillations} \label{sec:blochoscillations}
We consider loading noninteracting ultracold fermionic atoms in the lowest band of an optical lattice and applying a gradient potential to induce Bloch oscillations of them. This can be done by gravity, accelerating the optical lattice or applying a gradient magnetic field~\cite{anderson1998macroscopic,dahan1996bloch,tarruell2012creating}. At zero temperature and weak enough gradient potential, all the atoms populate the lowest band during the whole oscillation period. In ultracold-atom experiments, the center-of-mass velocity of atoms can be easily measured, given by ($\hbar=1$)
\begin{equation} \label{eq:com-velocity}
\bar{{\bf v}}(t)=\frac{1}{N}\int_{\mathrm{BZ}}\frac{d^{d}k}{(2\pi)^{d}}f^{p}({\bf k},t){\bf v}({\bf k}),
\end{equation}
where $N$ is the total number of particles, $d$ is the spatial dimension, $f^{p}({\bf k},t)$ is the nonequilibrium distribution function of fermionic atoms during the oscillations, and ${\bf v}({\bf k})$ is the atom velocity at the value of quasimomentum ${\bf k}$ in the lowest band, respectively. The integral in Eq.~\eqref{eq:com-velocity} is taken over the first Brillouin zone. In the limit of a weak and slowly varying gradient potential, a semiclassical approximation can be used to describe the drifts of atoms during the Bloch oscillations~\cite{chang1995berry,chang1996berry,sundaram1999wave,pettini2011anomalous,price2012mapping,dauphin2013extracting,jotzu2014experimental,aidelsburger2015measuring,spurrier2018semiclassical}. It results in a set of equations of motion in the following form,
\begin{subequations}  \label{eq:semi-classical}
\begin{eqnarray}
\mathbf{v({k})} &=& \dot{{\bf r}}=\nabla_{\bf k}\epsilon({\bf k})-\dot{\bf k}\times{\bf \Omega}({\bf k}),  \label{eq:semi-classical-1} \\
\dot{{\bf k}}(t) &=&{\bf F}(t),  \label{eq:semi-classical-2}
\end{eqnarray}
\end{subequations}
where $\epsilon({\bf k})$ is the energy dispersion of the lowest band, its derivative giving the group velocity, and ${\bf \Omega}({\bf k})$ is the Berry curvature of the lowest band, which contributes to the anomalous part of the velocity.

\subsection{Short-time dc driving}
We consider first the case of a time-independent driving force. In this situation, the solution of Eq.~\eqref{eq:semi-classical-2} is ${\bf k}(t)={\bf k}(0)+{\bf F}t$. We notice that the momentum of the atom drifts in the Brillouin zone at a constant speed. As a consequence, the distribution function evolves as $f^{p}({\bf k},t)=f_{0}^{p}({\bf k}-{\bf F}t)$, where $f_{0}^{p}({\bf k})$ is the equilibrium distribution function of atoms at $t=0$. Thus, the center-of-mass velocity of atoms during the Bloch oscillations can be obtained as $\bar{\bf v}(t)=\frac{1}{N}\int \frac{d^{d}k}{(2\pi)^{d}}f_{0}^{p}({\bf k}-{\bf F}t)\left\{ \nabla_{\bf k}\epsilon -{\bf F}\times{\bf \Omega}({\bf k})\right\} $. In this paper, we focus on the Hall response, i.e., the transverse velocity of an atomic cloud perpendicular to the driving force. This velocity is given by
\begin{equation} \label{eq:vx-dc}
\bar{\bf v}^{\mathrm{Hall}}(t) = -\frac{1}{N}\int \frac{d^{d}k}{(2\pi)^{d}}f_{0}^{p}({\bf k}-{\bf F}t){\bf F}\times{\bf \Omega}({\bf k}).
\end{equation}

In the short-time limit, the distribution function can be expanded up to the linear order of time as $f_{0}^{p}({\bf k}-{\bf F}t)=f_{0}^{p}({\bf k})-\nabla_{{\bf k}}f_{0}^{p}\cdot{\bf F}t+\cdots $. The mean velocity becomes therefore
\begin{equation} \label{eq:trans-mean-velocity}
\bar{v}_{\alpha}^{\mathrm{Hall}}(t) = \frac{1}{N}\left( \varepsilon_{\alpha
\beta \gamma}F_{\beta}D_{\gamma \eta}^{(0)}+\varepsilon_{\alpha \beta
\gamma}F_{\beta}D_{\gamma \eta}^{(1)}F_{\eta}t +\cdots \right),
\end{equation}
where $\varepsilon_{\alpha
\beta \gamma}$ represents the Levi-Civita tensor and $D_{\gamma \eta}^{(n)}=\int \frac{d^{d}k}{(2\pi)^{d}} \Omega_{\gamma}({\bf k})\frac{\partial^{n}f_{0}^{p}}{\partial k_{\eta}^{n}}=(-1)^{n}\int \frac{d^{d}k}{(2\pi)^{d}} \frac{\partial^{n}\Omega_{\gamma}}{\partial k_{\eta}^{n}}f_{0}^{p}({\bf k})$. Here, $D_{\gamma \eta}^{(0)}$ is the coefficient of the linear Hall response and $D_{\gamma \eta}^{(1)}$ is the so-called Berry curvature dipole (BCD) tensor, which is the second-order coefficient of the nonlinear Hall response, respectively. Looking at the oscillation dynamics of ultracold fermions, this nonlinear Hall response can be determined. More specifically, by measuring the early-stage growth rate of transverse Hall velocity, we can obtain the BCD tensor, which reflects the distribution of the Berry curvature in the Brillouin zone.

For the bands with inversion symmetry ${\bf \Omega}(-{\bf k})={\bf \Omega}({\bf k})$, while ${\bf \Omega}(-{\bf k})=-{\bf \Omega}({\bf k})$ for the systems with time-reversal symmetry. The presence of both IS and TRS means zero Berry curvature in the whole Brillouin zone, ${\bf \Omega}({\bf k})=0$. As a result, the transverse current vanishes at arbitrary order. If TRS is broken, there is a linear-order Hall response, i.e., $D_{\gamma \eta}^{(0)}\neq 0$. When the system has TRS but breaks IS, the linear Hall response vanishes, $D_{\gamma \eta}^{(0)}=0$, since $f_{0}^{p}({\bf k})\Omega_{\gamma}({\bf k})$ is an odd function in the Brillouin zone. However, the second-order nonlinear Hall response may be manifested, $D_{\gamma \eta}^{(1)}\neq 0$, since $\frac{\partial \Omega_{\gamma}}{\partial k_{\eta}}f_{0}^{p}({\bf k})$ becomes an even function.
\begin{figure}[t]
\centering
\begin{tikzpicture}[>=latex,scale=0.8] 
    \coordinate (Origin)   at (0,0);
    \coordinate (Borigin)  at (0.5,-{sin{60}});
    \coordinate (XAxisMin) at (-1.5,0);
    \coordinate (XAxisMax) at (1.5,0);
    \coordinate (YAxisMin) at (0,-0.5);
    \coordinate (YAxisMax) at (0,2.6);
    \draw [thick,dashed,gray,-latex] (XAxisMin) -- (XAxisMax);
    \draw [thick,dashed,gray,-latex] (YAxisMin) -- (YAxisMax);
    \node[draw=none,above,gray] at (XAxisMax){$x$};
    \node[draw=none,left,gray] at (YAxisMax){$y$};

    \clip (-2,-2) rectangle (4cm,3cm); 
    \coordinate (Aone) at (0,{sqrt(3)});
    \coordinate (Atwo) at (1.5,-{sin{60}});
    \coordinate (Athree) at (-1.5,-{sin{60}});
    \coordinate (Bone) at (0.5,{sqrt(3)-sin{60}});
    \coordinate (Btwo) at (2,{-sqrt(3)});
    \coordinate (Bthree) at (-1.0,-{sqrt(3)});
    \coordinate (Done) at (-1,0);
    \coordinate (Dtwo) at (0.5,{sin{60}});
    \coordinate (Dthree) at (0.5,-{sin{60}});
    \node[draw=none,above right,red] at (Dtwo){${\bf A}$};
    \node[draw=none,above left,blue] at (1.5,sin{60}){${\bf B}$};
  \foreach \i in{-3,...,3}{
  \foreach \j in{-3,...,3}{
  \node[draw,circle,inner sep=1.5pt,fill,blue] at (3*\i,2*sin{60}*\j){};
  \node[draw,circle,inner sep=1.5pt,fill,blue] at (3*\i+3*cos{60},2*sin{60}*\j+sin{60}){};
  \node[draw,circle,inner sep=1.5pt,fill,red] at (3*\i+0.5,2*sin{60}*\j+sin{60}){};
  \node[draw,circle,inner sep=1.5pt,fill,red] at (3*\i-1,2*sin{60}*\j){};
  \foreach \a in{0,120,-120} \draw (3*\i,2*sin{60}*\j) -- +(\a:-1);
  \foreach \a in{0,120,-120} \draw (3*\i-3*cos{60},2*sin{60}*\j-sin{60}) -- +(\a:-1);
  }}
    \draw [ultra thick,-latex,blue] (Origin) -- (Aone) node [below left]{${\bf a}_1$};
    \draw [ultra thick,-latex,blue] (Origin) -- (Atwo) node [above]{${\bf a}_2$};
    \draw [ultra thick,-latex,blue] (Origin) -- (Athree) node [below right]{${\bf a}_3$};
    \draw [thick,-latex,green] (Origin) -- (Done) node [above right]{${\bf d}_1$};
    \draw [thick,-latex,green] (Origin) -- (Dtwo) node [below right]{${\bf d}_2$};
    \draw [thick,-latex,green] (Origin) -- (Dthree) node [left]{$d_3$};
    \draw[<->,thick,yshift=-0.1cm, decoration={markings,mark= at position 0.5 with{\node[below] at (0,0){$J_1$}; }}, postaction={decorate}] (2, 0) to[bend right] (3,0);
    \draw[<->,thick,scale=0.98,dashed,xshift=0.12cm, decoration={markings,mark= at position 0.5 with{\node[below right] at (0,0){$J^\prime_1$}; }}, postaction={decorate}] (3, 0) to[bend right] (3.5,sin{60});
    \draw[<->,thick,scale=0.98,dashed,xshift=0.12cm, decoration={markings,mark= at position 0.5 with{\node[below] at (0,0){}; }}, postaction={decorate}] (3, 0) to[bend left] (3.5,-sin{60});
   \draw[->,thick,dotted,red, decoration={markings,mark= at position 0.5 with{\node[right] at (0,0){$J_2e^{-i\phi}$}; }}, postaction={decorate}] (2,{-0.1}) to (2,{-2*sin{60}+0.1}); 
    \draw[->,thick,dotted,red, decoration={markings,mark= at position 0.5 with{\node[] at (0,0){}; }}, postaction={decorate}] (2,{-2*sin{60}}) to ({3+sin{60}/2-0.05},{-sin{60}-0.05});
    \draw[->,thick,dotted,red, decoration={markings,mark= at position 0.5 with{\node[] at (0,0){}; }}, postaction={decorate}] ({3+sin{60}/2},-sin{60}) to (2.1,{-0.05});
    \draw[->,thick,dashed,blue, decoration={markings,mark= at position 0.5 with{\node[left] at (0,0){$J_2e^{i\phi}$}; }}, postaction={decorate}] (3, 2*sin{60}) to (3,{0.1});
    \draw[->,thick,dashed,blue, decoration={markings,mark= at position 0.5 with{\node[] at (0,0){}; }}, postaction={decorate}] (3, 0) to ({1+sin{60}/2+0.15},{sin{60}-0.05});
    \draw[->,thick,dashed,blue, decoration={markings,mark= at position 0.5 with{\node[] at (0,0){}; }}, postaction={decorate}] ({1+sin{60}/2},{sin{60}}) to (2.9,{2*sin{60}-0.05});
    \end{tikzpicture}%
\begin{tikzpicture}[>=latex,scale=0.6] 
    \coordinate (Origin)   at (0,0);
    \coordinate (Borigin)  at (0.5,-{sin{60}});
    \coordinate (XAxisMin) at (-2,0);
    \coordinate (XAxisMax) at (2,0);
    \coordinate (YAxisMin) at (0,-2);
    \coordinate (YAxisMax) at (0,2);

    \clip (-3,-2.5) rectangle (3cm,3cm); 
    \coordinate (Bone) at ({pi/3},{pi/2/sin{60}});
    \coordinate (Btwo) at ({2*pi/3},0);
   \newdimen\RR
   \RR=1cm
   \draw (30:\RR) \foreach \x in{90,150,...,390}{  -- (\x:\RR) } [fill=gray];
   \foreach \x/\l/\p in
    { 90/{$K$}/above,
      150/{$K^\prime$}/above,
      210/{$K$}/left,
      270/{$K^\prime$}/below,
      330/{$K$}/below,
      390/{$K^\prime$}/right
      }
   \node[inner sep=0pt,circle,draw,fill,label={\p:\l}] at (\x:\RR){};
   \newdimen\R
   \R=2cm
   \draw (0:\R) \foreach \x in{60,120,...,360}{  -- (\x:\R) };
   \foreach \x/\l/\p in
    { 60/{$\Gamma$}/above,
      120/{$\Gamma$}/above,
      180/{$\Gamma$}/left,
      240/{$\Gamma$}/below,
      300/{$\Gamma$}/below,
      360/{$\Gamma$}/right }
     \node[inner sep=1.5pt,circle,draw,fill,label={\p:\l}] at (\x:\R){};
     \node[inner sep=1.5pt,circle,draw,fill,label={$\Gamma$}] at (Origin){};
    \draw [thick,-latex,red] (Origin) -- (Bone) node [below right]{${\bf b}_1$};
    \draw [thick,-latex,red] (Origin) -- (Btwo) node [below left]{${\bf b}_2$};
    \end{tikzpicture}
\caption{The Haldane model and associated Brillouin zone. $J_1$ ($J_1^{\prime}$) and $J_2$ denote the nearest-neighbor and the next-to-nearest-neighbor hopping amplitudes, respectively. $\phi$ is the phase of the next-to-nearest-neighbor hopping.}
\label{fig:honeycomb}
\end{figure}
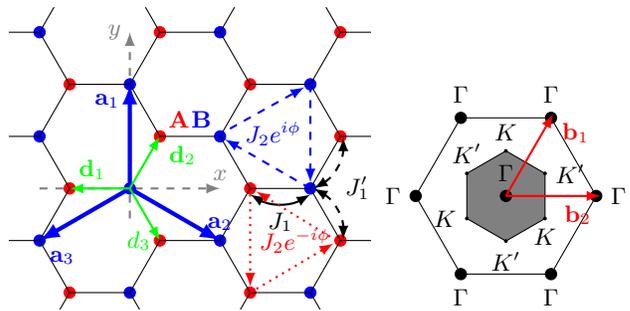

To illustrate the mentioned nonlinear Hall response and its dependence on band symmetry, we numerically investigate Bloch oscillations in the Haldane model~\cite{haldane1988model} (see Fig.~\ref{fig:honeycomb}). This model was realized by circularly shaking a two-dimensional honeycomb optical lattice in an ultracold-atom system~\cite{oka2009photovoltaic,zheng2014floquet,jotzu2014experimental}. The Hamiltonian for this system can be written as $H({\bf k})=E_{0}({\bf k})\cdot \mathbb{I}+{\bf B}({\bf k})\cdot \boldsymbol{\sigma}$, where $\mathbb{I}$ is the $2\times 2$ identity matrix and $\boldsymbol{\sigma}=(\sigma_{x},\sigma_{y},\sigma_{z})$ are the Pauli matrices. The elements of this Hamiltonian are expressed as follows:
\begin{subequations}
\begin{eqnarray}
E_{0}({\bf k}) &=& 2J_{2}\cos{\phi}\sum_{\alpha =1,2,3}\cos{({\bf k\cdot a}_{\alpha})}, \\
B_{x}({\bf k}) &=& -J_{1}\cos{({\bf k\cdot d}_{1})}-J_{1}^{\prime}\sum_{\alpha =2,3}\cos{({\bf k\cdot d}_{\alpha})}, \\
B_{y}({\bf k}) &=& -J_{1}\sin{({\bf k\cdot d}_{1})}-J_{1}^{\prime}\sum_{\alpha =2,3}\sin{({\bf k\cdot d}_{\alpha})}, \\
B_{z}({\bf k}) &=& M+2J_{2}\sin{\phi}\sum_{\alpha}\sin{({\bf k\cdot
a}_{\alpha})}.
\end{eqnarray}
\end{subequations}
Here, $J_{1}$ ($J_{1}^{\prime}$) and $J_{2}$ denote the nearest-neighbour and the next-to-nearest-neighbour hopping strength, $\phi$ is the phase of the next-to-nearest-neighbour hopping that breaks the TRS, and $M$ is the imbalance of the $AB$ sublattice that breaks the IS, respectively. The mentioned parameters can be well controlled in current experiments with cold atoms~\cite{tarruell2012creating,jotzu2014experimental}.  In this two-dimensional model, the Berry curvature is restricted along the $z$ axis, i.e., ${\bf \Omega}({\bf k})= {\bf e}_{z}\Omega_{z}({\bf k})$. Setting the value $\phi=0$ leads to the restoration of TRS. In this case, Eq.~\eqref{eq:trans-mean-velocity} can be expanded as follows:
\begin{subequations} \label{eq:J-currents}
\begin{eqnarray}
\bar{v}_{x}^{\mathrm{Hall}}(t) &=&\frac{1}{N}\left(F_{y}D_{zx}^{(1)}F_{x}t+F_{y}D_{zy}^{(1)}F_{y}t\right), \\
\bar{v}_{y}^{\mathrm{Hall}}(t) &=&\frac{1}{N}\left(-F_{x}D_{zx}^{(1)}F_{x}t-F_{x}D_{zy}^{(1)}F_{y}t\right),
\end{eqnarray}
\end{subequations}
In the coordinate system shown in Fig.~\ref{fig:honeycomb}, the lattice has a reflection symmetry in the $y$ direction, but breaks it along the $x$ axis. This results in $\Omega_{z}(-k_{x},k_{y})=\Omega_{z}(k_{x},k_{y})$. The TRS ensures $\Omega_{z}(-{\bf k})=-\Omega_{z}({\bf k})$, giving $\Omega_{z}(k_{x},-k_{y})=-\Omega_{z}(k_{x},k_{y})$. In this case, $\partial \Omega_{z}/\partial k_{y}$ is an even function in the Brillouin zone, while $\partial \Omega_{z}/\partial k_{x}$ is odd. Therefore, $D_{zx}^{(1)}$ vanishes but $D_{zy}^{(1)}$ remains. From now on, we will set an external force only along the $y$ direction so that Eqs.~\eqref{eq:J-currents} are simplified as $\bar{v}_{y}^{\mathrm{Hall}}(t)=0$ and $\bar{v}_{x}^{\mathrm{Hall}}(t)=\frac{1}{N}D_{zy}^{(1)}F_{y}^{2}t$.
\begin{figure}[t]
\centering
\includegraphics[width=0.48\textwidth]{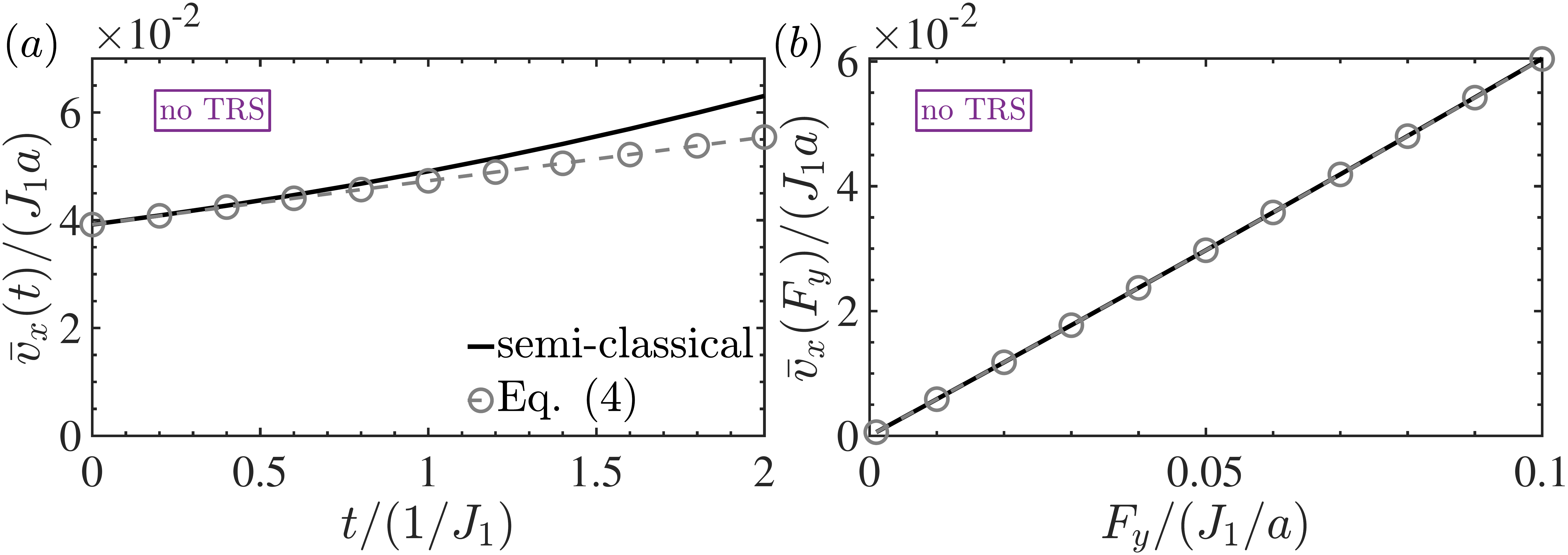}\\
\includegraphics[width=0.48\textwidth]{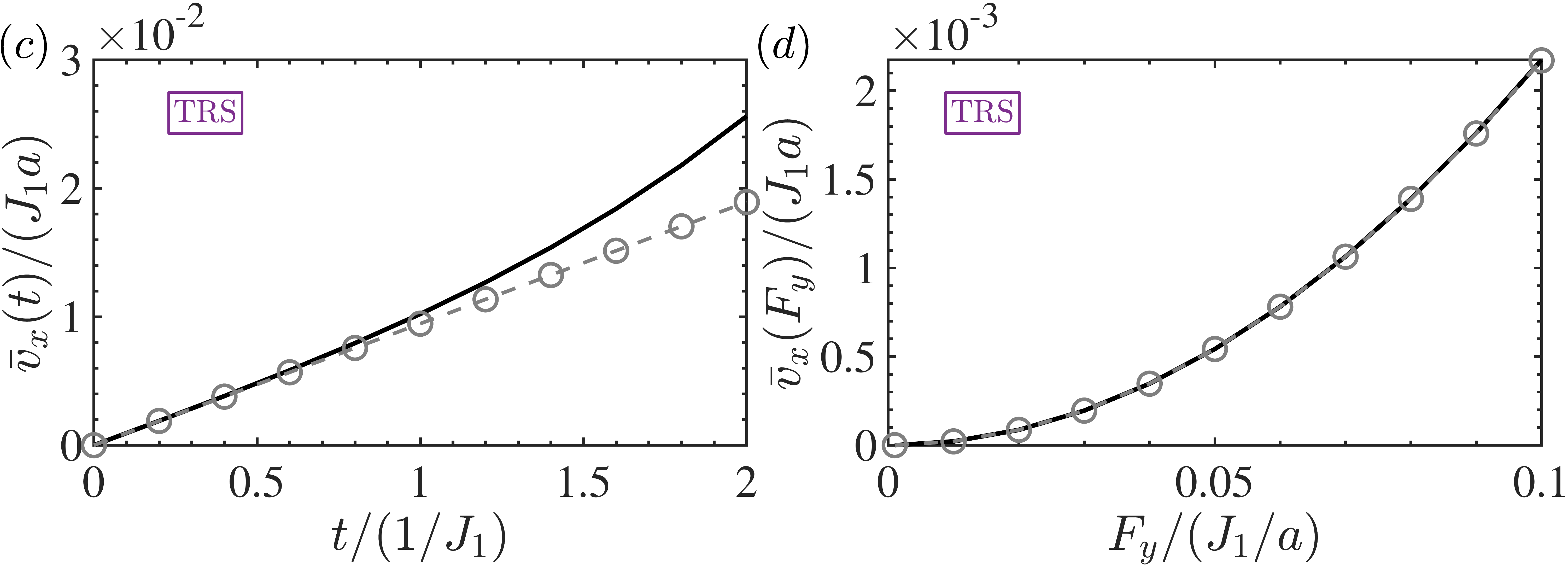}
\caption{Mean velocity $\bar{v}_{x}$ calculated from semiclassical dynamics (solid lines) and by Eq.~\eqref{eq:trans-mean-velocity} (gray-dashed lines with circles): (a), (c) Over a small time interval $t$ at a small value of $F_y$, and (b), (d) as functions of $F_y$ at a definite short-time value. The hopping ratios are set to $J^\prime_1/J_1=0.6$, $J_2/J_1=0.1$, and the mass term is $M/J_1=-0.2$ breaking the inversion symmetry. $\phi$ is set to be $\pi/8$ (or $0$) in (a), (b) [or (c), (d)] to break (or preserve) the time-reversal symmetry.}
\label{fig2}
\end{figure}

In Fig.~\ref{fig2}, the mean Hall velocities $\bar{v}_{x}$ calculated by the semiclassical equations of motion~\eqref{eq:semi-classical} and by the short-time series expansion~\eqref{eq:trans-mean-velocity} are presented. In both cases with and without TRS provided in Figs.~\ref{fig2}(a) and~\ref{fig2}(c), the results obtained by Eq.~\eqref{eq:trans-mean-velocity} (gray-dashed lines with circles) overlap with the Hall velocity $\bar{v}_{x}$ data (solid lines) obtained from semiclassical dynamics in a short-time interval. In the presence of TRS, the nonlinear contribution plays a dominant role at short times and $\bar{v}_{x}$ exhibits a parabolic dependence on $F_{y}$ as shown in Fig.~\ref{fig2}(d). When TRS is broken as in Fig.~\ref{fig2}(b), the linear part is dominant at short times and $\bar{v}_{x}$ becomes a linear dependence on $F_{y}$.

\begin{figure}[t]
\centering
\includegraphics[width=0.48\textwidth]{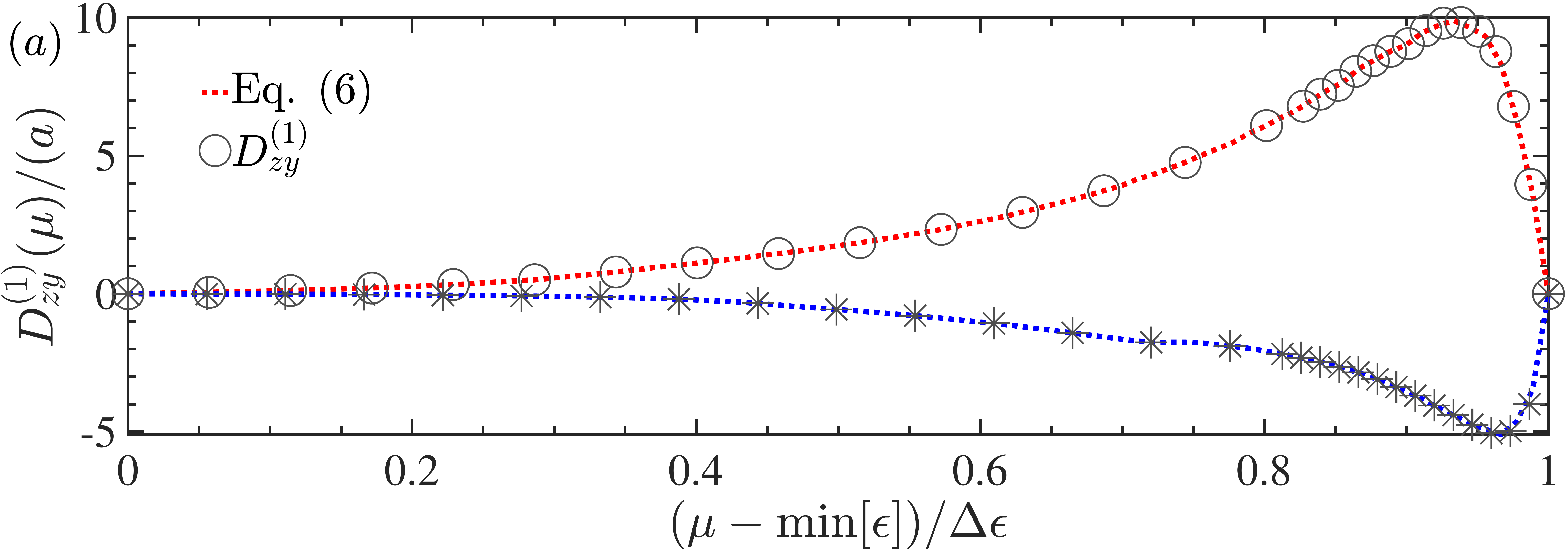}\\
\includegraphics[width=0.48\textwidth]{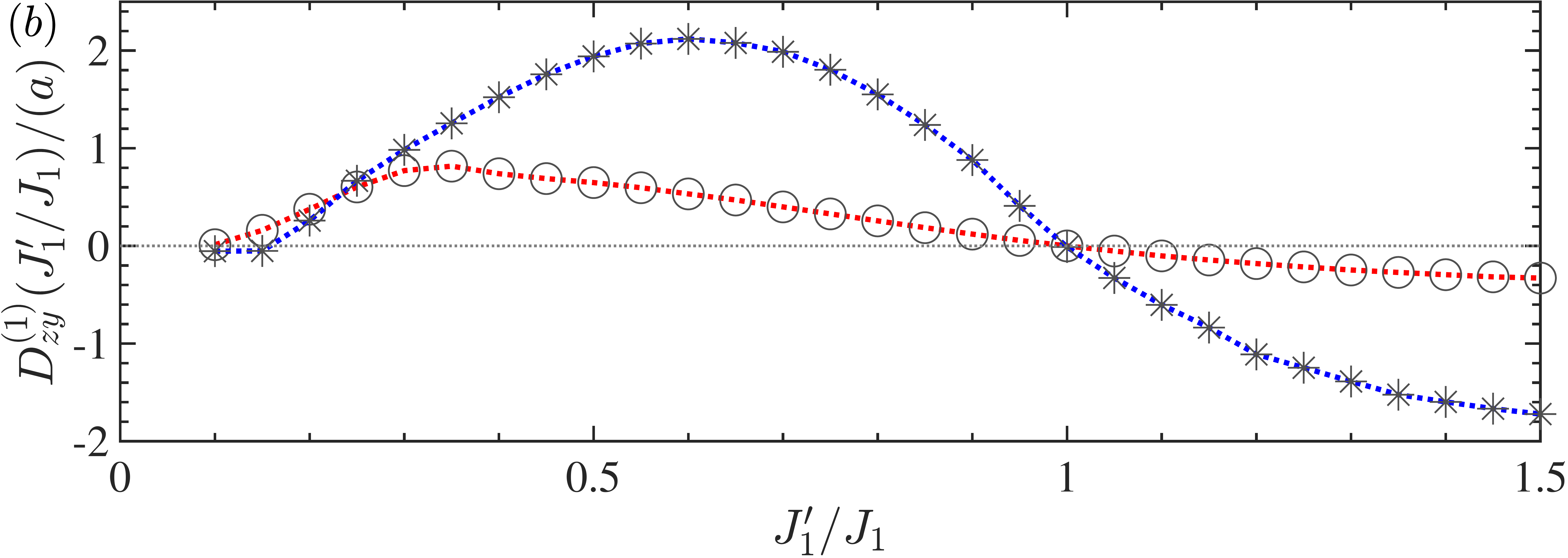}
\caption{Berry curvature dipole $D^{(1)}_{zy}$ calculated by the analytic expression (symbols) and using the values of $\bar{v}_x$ found by Eq.~\eqref{eq:J-currents} (lines) as a function of (a) filling of the lowest band $(\mu-\mathrm{min}[\epsilon])/\Delta\epsilon$ at $J^\prime_1/J_1=0.6$, $1.4$, and (b) hopping ratio $J^\prime_1/J_1$ at two values of the lowest band filling, i.e., 0.5 and 0.8. Other parameters are the same as in Fig.~\ref{fig2}.} 
\label{fig3}
\end{figure}

We calculated the BCD tensor of the honeycomb lattice at different values of $J_{1}^{\prime}/J_{1}$ and chemical potential $\mu$, as shown in Fig.~\ref{fig3}. We notice that the components of the BCD tensor become zero at $J_{1}^{\prime}=J_{1}$ for any given filling of fermions because of the $C_{3}$ symmetry of the system. After a rotation of the lattice by the angle $\theta$ in the $xy$ plane, the Berry curvature $\tilde{\Omega}_{z}^{(\theta)}(k_{x},k_{y})$ becomes
\begin{equation*}
\tilde{\Omega}_{z}^{(\theta)}(k_{x},k_{y}) = \Omega_{z}(k_{x}\cos{\theta} +k_{y}\sin{\theta},-k_{x}\sin{\theta}+k_{y}\cos{\theta}).
\end{equation*}
Therefore, the BCD tensor after rotation reads
\begin{eqnarray*}
\tilde{D}_{zy}^{(1)} &=& \int_{\mathrm{BZ}}\frac{d^{2}k}{(2\pi)^{2}}\frac{\partial \tilde{\Omega}_{z}^{(\theta)}}{\partial k_{y}}f_{0}^{p}({\bf k},\mu) 
\\
&=&\int_{\mathrm{BZ}}\frac{d^{2}k}{(2\pi)^{2}}\left( \frac{\partial \Omega_{z}}{\partial k_{x}}\sin{\theta}+\frac{\partial \Omega_{z}}{\partial k_{y}}\cos{\theta}\right) f_{0}^{p}({\bf k},\mu) 
\\
&=& D_{zx}^{(1)}\sin{\theta}+D_{zy}^{(1)}\cos{\theta},
\end{eqnarray*}
where $D_{zx}^{(1)}=0$. If the system has a $C_{3}$ symmetry, rotation by $\theta =2\pi /3$ will keep it invariant. Therefore, $\tilde{\Omega}_{z}^{(2\pi /3)}=\Omega_{z}$ and $\tilde{D}_{zy}^{(1)}=D_{zy}^{(1)}$, thus making $D_{zy}^{(1)}\cos{\theta}=D_{zy}^{(1)}$ and, hence, $D_{zy}^{(1)}=0$. This conclusion can be generalized by stating that any system with discrete rotation symmetry has a vanishing BCD tensor.

The BCD tensor as a function of the chemical potential is presented in Fig.~\ref{fig3}(a). When the chemical potential increases from the bottom of the energy band, the BCD tensor increases first with the chemical potential, reaching its maximum. It decreases then to zero at half filling when the chemical potential exceeds the top boundary of the lower band. This phenomenon can be understood in two limits, i.e., at low and high fillings. In the low filling limit, the tight-binding Hamiltonian can be expanded near the bottom of the lower band, i.e., around $\mathbf{k=0}$ as follows,
\begin{eqnarray*}
H(\mathbf{k)} &\simeq & M\sigma_{z} +(\lambda -1)J_{1}^{\prime}a k_{x}\sigma_{y}
\\
&& +\left\{ -(\lambda +2)J_{1}^{\prime}+ \frac{J_{1}^{\prime}a^2}{4}\left[ (2\lambda +1)k_{x}^{2}+3k_{y}^{2}\right]
\right\} \sigma_{x}, 
\end{eqnarray*}
with $J_{1}\equiv \lambda J_{1}^{\prime}$, and $a$ being the lattice constant, respectively. The associated Berry curvature is expressed as follows:
\begin{eqnarray}
\Omega_{z}({\bf k}) &=& \frac{M}{2|{\bf B(k)}|^{3}}\left( \frac{\partial B_{x}}{\partial k_{x}}\frac{\partial B_{y}}{\partial k_{y}}-\frac{\partial B_{y}}{\partial k_{x}}\frac{\partial B_{x}}{\partial k_{y}}\right) \nonumber
\\
&=&\frac{3M}{4|{\bf B(0)}|^{3}}\left( 1-\lambda \right){J_{1}^{\prime}}^{2}a^3k_{y}.
\end{eqnarray}
The function $\partial \Omega_{z}/\partial k_{y}$ becomes a constant at ${\bf k\approx0}$. Therefore, the BCD tensor is proportional to the volume $A_{p}(\mu)$ of the Fermi sea, $D_{zy}^{(1)}\propto A_{p}(\mu)$. With the increase of $\mu$, the volume of the particle Fermi sea grows from zero, making the BCD tensor increase with the chemical potential. In addition, depending on the sign of $1-\lambda=1-J_{1}/J_{1}^{\prime}$, the BCD tensor will change its sign as shown in Fig.~\ref{fig3}.

At high fillings, the Hamiltonian near the top boundary of the energy band, i.e., around two Dirac points, can be expanded as follows,
\begin{eqnarray*}
H^{(1)},H^{(2)} &\simeq & M\sigma_{z}\pm \left( \eta v_{_F}p_{y}\pm \frac{\lambda a v_{_F}}{4}(p_{x}^{2}-p_{y}^{2})\right) \sigma_{x} 
\\
&& +\left( \lambda v_{_F}p_{x}\pm \frac{\eta a v_{_F}}{2}p_{x}p_{y}\right) \sigma_{y},
\end{eqnarray*}
where $\eta =\sqrt{(4-\lambda^{2})/3}$, $v_{_F}=-3J_{1}^{\prime}a/2$, ${\bf p=k-k}_{D}$, and ${\bf k}_{D}$ defined the positions of the Dirac points. The Berry curvatures near two Dirac points are
\begin{equation*}
\Omega_{z}^{(1)},\Omega_{z}^{(2)}\simeq \mp \frac{\lambda \eta M v_{F}^{2}}{2|M|^{3}}
\left( 1 \mp \frac{\chi a p_{y}}{2}\right),
\end{equation*}
where $\chi=\lambda /\eta -\eta /\lambda$. We obtain therefore
\begin{equation}
\frac{\partial \Omega_{z}^{(1)}}{\partial k_{y}},\frac{\partial \Omega_{z}^{(2)}}{\partial k_{y}}\simeq \frac{\lambda \eta M v_{F}^{2} a }{4|M|^{3}}\chi.
\end{equation}
Notice that $\partial \Omega_{z}/\partial k_{y}$ is also a constant near the Dirac points in the high filling limit. The Berry curvature dipole can be written as follows,
\begin{eqnarray*}
D_{zy}^{(1)} &=& \int_{\mathrm{BZ}}\frac{d^{2}k}{(2\pi)^{2}} \frac{\partial \Omega_{z}}{\partial k_{y}}f_{0}^{p}({\bf k},\mu) 
\\
&=& -\int_{\mathrm{BZ}}\frac{d^{2}k}{(2\pi)^{2}}\frac{\partial \Omega_{z}}{\partial k_{y}}f_{0}^{h}({\bf k},\mu),
\end{eqnarray*}
where $f_{0}^{h}({\bf k},\mu)$ is the equilibrium distribution function of holes. Here, we have used the fact that $\int_{\mathrm{BZ}}\frac{d^{2}k}{(2\pi)^{2}}\frac{\partial \Omega_{z}}{\partial k_{y}}=0$. At high fillings, there are two Fermi surfaces of holes near two Dirac points, $f_{0}^{h}({\bf k},\mu)=f_{0}^{h(1)}({\bf k},\mu )+f_{0}^{h(2)}({\bf k},\mu)$. Hence,
\begin{equation*}
D_{zy}^{(1)} = -\int_{\mathrm{BZ}}\frac{d^{2}k}{(2\pi)^{2}}\left[ \frac{\partial \Omega_{z}^{(1)}}{\partial k_{y}}f_{0}^{h(1)}({\bf k},\mu)+\frac{\partial \Omega_{z}^{(2)}}{\partial k_{y}}f_{0}^{h(2)}({\bf k},\mu)\right].
\end{equation*}
After integration, we obtain $D_{zy}^{(1)}\propto A_{h}(\mu)$, where $A_{h}(\mu)$ is the volume of the Fermi sea of holes. The latter decreases when $\mu$ approaches to the band top, and the BCD tensor decreases towards zero.

\subsection{Long-time dc driving}
\begin{figure}[t]
\centering
\includegraphics[width=0.48\textwidth]{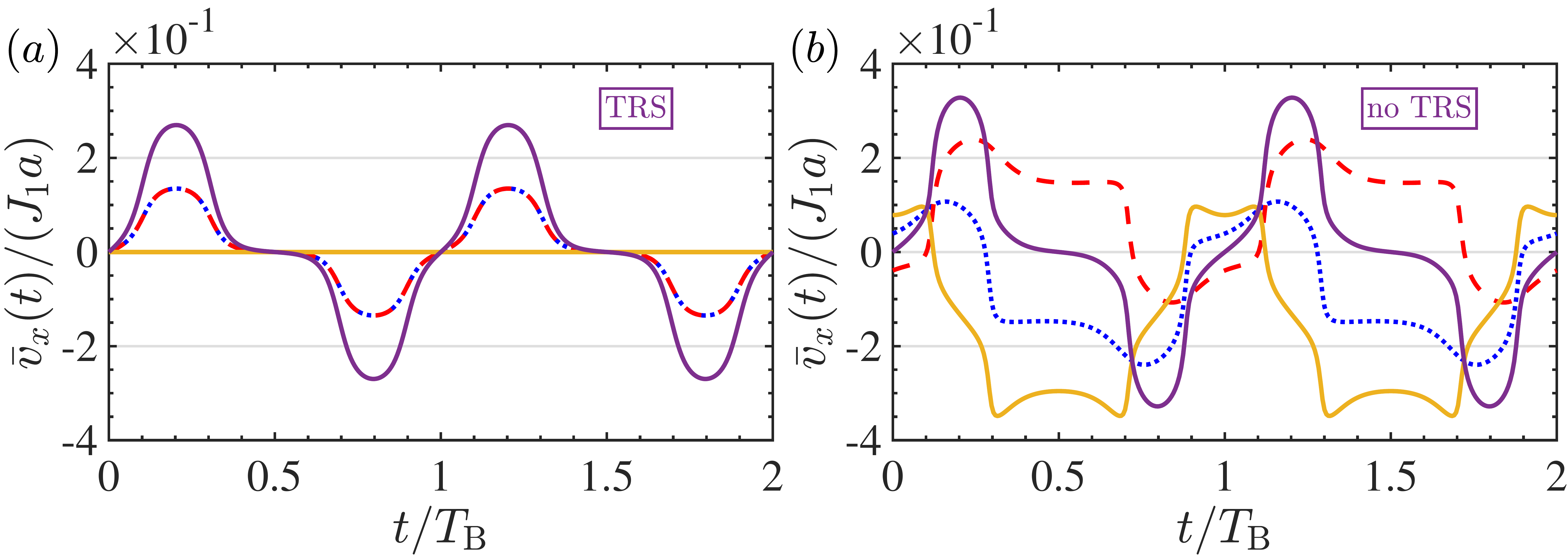}\\
\includegraphics[width=0.48\textwidth]{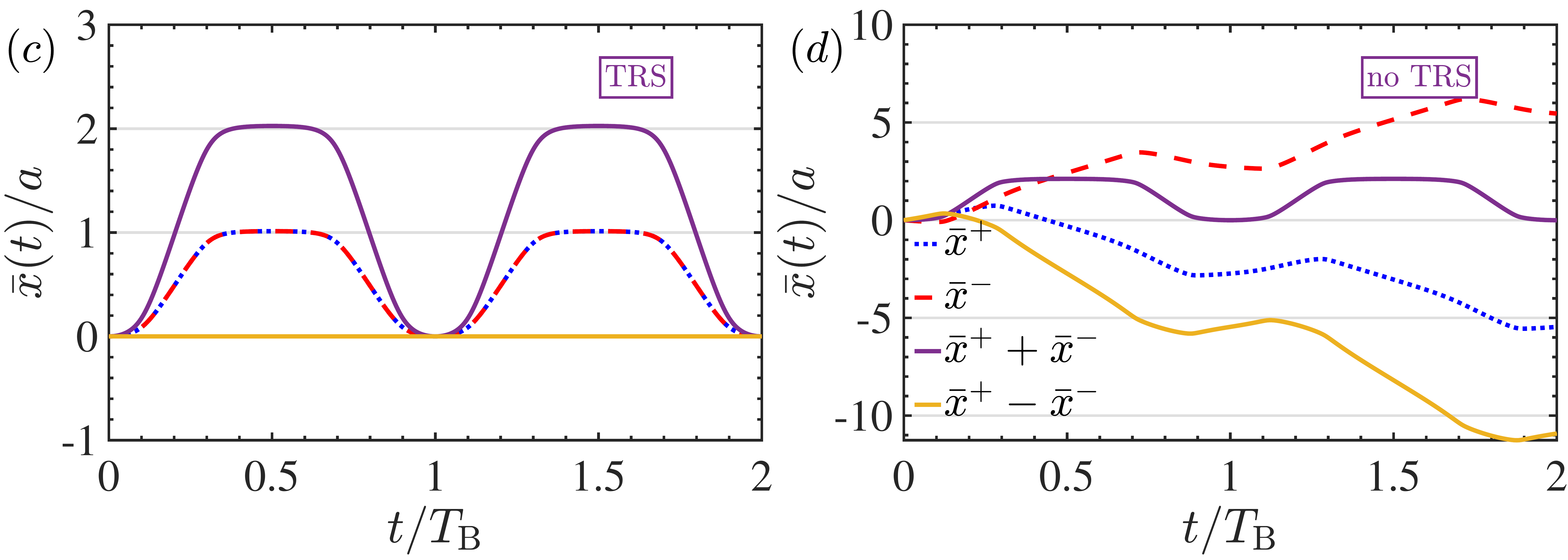}
\caption{Time-dependent mean velocity $\bar{v}_x(t)$ and displacement $\bar{x}(t)$ during two Bloch oscillation periods $2T_\mathrm{B}$ for opposite constant forces $\pm F_y$ with and without TRS. The individual velocities $\bar{v}_x^{\pm}(t)$ or displacements $\bar{x}^{\pm}(t)$ for opposite directions are denoted in dotted-blue and dashed-red lines,  and their sum and difference are shown in purple and yellow solid lines, respectively. Other parameters are the same as in Fig.~\ref{fig2}.}
\label{fig4}
\end{figure}
We turn to consider the long-time behavior of the center-of-mass displacement or drift for an entire Bloch oscillation as measured in the experiment in Ref.~\citep{jotzu2014experimental}. In the presence of the $y$-axis external dc force $F_y$, at a time $t$, the time-dependent transverse displacement for atoms initially at ($k_x^0,k_y^0$) is given as
\begin{eqnarray*}
x(t) &=&\int_0^t  v_x(k_x,k_y(t'))dt' \\
&=& \frac{1}{F_y}\int_{k^0_y}^{k^0_y+F_yt}\left[\partial_{k^0_x} \epsilon (k^0_x,k_y) - F_y \Omega(k^0_x,k_y)\right] dk_y  \\
&\equiv& \mathcal{A}(k^0_x,k^0_y,F_yt)+\mathcal{B}(k^0_x,k^0_y,F_yt). 
\end{eqnarray*}
In the presence of TRS, we have $\partial_{k_a}\epsilon({\bf k})=-\partial_{k_a}\epsilon(-{\bf k})$ and $\Omega({\bf k})=\Omega(-{\bf k})$ such that two parts in $x(t)$ can be rewritten as
\begin{eqnarray*}
\mathcal{A}(-k^0_x,-k^0_y,F_yt) &=& \frac{1}{F_y}\int_{-k^0_y}^{-k^0_y+F_yt} \partial_{-k^0_x} \epsilon (-k^0_x,k_y) dk_y \\
&=& \frac{1}{F_y}\int_{k^0_y}^{k^0_y-F_yt} \partial_{-k^0_x} \epsilon (-k^0_x,-k_y) d(-k_y) \\
&=& -\frac{1}{F_y}\int_{k^0_y-F_yt}^{k^0_y} \partial_{k^0_x} \epsilon (k^0_x,k_y) dk_y,
\\
\mathcal{B}(-k^0_x,-k^0_y,F_yt) &=& -\int_{-k^0_y}^{-k^0_y+F_yt} \Omega(-k^0_x,k_y) dk_y \\
&=& -\int_{k^0_y}^{k^0_y-F_yt} \Omega(-k^0_x,-k_y) d(-k_y) \\
&=& \int_{k^0_y-F_yt}^{k^0_y} \Omega(k^0_x,k_y) dk_y. 
\end{eqnarray*}
It is straightforward to see that, after an entire Bloch circle $T_\mathrm{B}$ with the equivalence between intervals $[k^0_y,k^0_y+F_yT_\mathrm{B}]$ and $[k^0_y-F_yT_\mathrm{B},k^0_y]$, we obtain two odd functions $\mathcal{A}(-k^0_x,-k^0_y,F_yT_\mathrm{B})=-\mathcal{A}(k^0_x,k^0_y,F_yT_\mathrm{B})$ and $\mathcal{B}(-k^0_x,-k^0_y,F_yT_\mathrm{B})=-\mathcal{B}(k^0_x,k^0_y,F_yT_\mathrm{B})$ in the Brillouin zone. Therefore, after integrating over the Brillouin zone with a distribution, the drift $\bar{x}(t)$ vanishes after a Bloch period $t=T_\mathrm{B}$. In contrast, these two functions are not guaranteed to be odd when breaking TRS and the drift will be nonzero after a Bloch period. These results are verified by our numerical calculations as shown in Fig.~\ref{fig4}. The nonzero drift without TRS after a Bloch period has already been justified from the measurement in Ref.~\cite{jotzu2014experimental}.

\begin{figure}[t]
\centering
\includegraphics[width=0.48\textwidth]{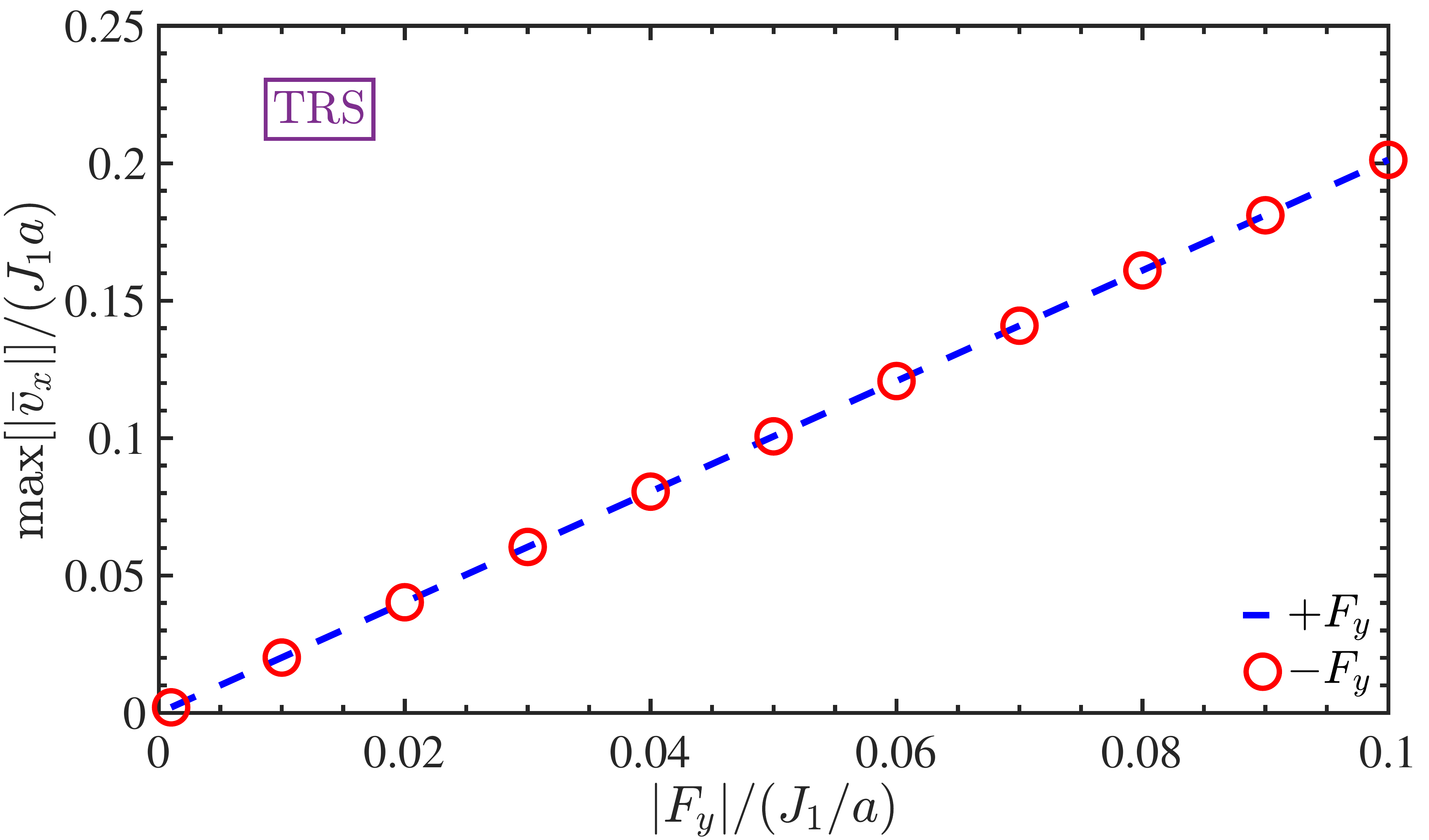}
\caption{Maximum of the mean velocity $\bar{v}_x(t)$ in one Bloch oscillation period, as functions of the magnitude of forces $\pm F_y$ with TRS. Other parameters are the same as in Fig.~\ref{fig2}.}
\label{fig5}
\end{figure}

In addition, we discuss the relation between the maximum value of $\bar{v}_x(t)$ in one Bloch period and the external force $F_y$. In the presence of a $y$-axis force $F_y$, the time-dependent transverse velocity in Eq.~\eqref{eq:vx-dc} can be rewritten as
\begin{equation*}
    \bar{v}_x(t) = -\frac{F_y}{N}\int \frac{d^{2}k}{(2\pi)^{2}}f_{0}^{p}({\bf k}) \Omega_z({\bf k}+{\bf e}_y F_yt).
\end{equation*}
By introducing $\delta k=F_yt$, the derivative of velocity can be written as
\begin{equation*}
\frac{\partial \bar{v}_x}{\partial t}=\frac{\partial \bar{v}_x}{\partial (\delta k)}F_y=-\frac{F_y^2}{N}\int \frac{d^{2}k}{(2\pi)^{2}}f_{0}^{p}({\bf k}) \frac{\partial \Omega_z({\bf k}+{\bf e}_y \delta k)}{\partial k_y}.
\end{equation*}
Then, by solving $\partial \bar{v}_x/\partial (\delta k)=0$, one can find the extremum $\delta k_m$ for the maximum of $\bar{v}_x$. It is straightforward to see that $\delta k_m$ becomes independent of $F_y$ and $t$. Thus, the maximum transverse velocity in one Bloch period is given by
\begin{eqnarray*}
\mathrm{max}[\bar{v}_x(t)] &=& -\frac{F_y}{N}\int \frac{d^{2}k}{(2\pi)^{2}}f_{0}^{p}({\bf k}) \Omega_z({\bf k}+{\bf e}_y \delta k_m) \\
&=& -\frac{F_y}{N}\left\{\sum_n\frac{1}{n!}D^{(n)}_{zy}(\delta k_m)^n\right\}.
\end{eqnarray*}
In the presence of TRS as in this work, the linear-response term with $D^{(0)}_{zy}$ vanishes. However, we can see that the nonlinear contributions with $D^{(n)}_{zy}$ ($n\geq1$) remain nonzero, leading to a linear dependence of the maximum velocity on the magnitude of external force. This result is also verified by our numerical calculations which are shown in Fig.~\ref{fig5}.

\subsection{ac driving}
In this section, we further consider driving the atoms by an oscillating force, ${\bf F}(t)=F_{y}{\bf e}_{y}\cos{(\omega t)}$. The driving frequency $\omega$ is set to be much smaller than the band gap, so that the interband transitions are highly suppressed, and the atoms remain in the lowest band during the Bloch oscillations. Therefore, the quasimomentum of atoms in the lowest band is given by ${\bf k}(t)={\bf k}(0)-\frac{F_{y}{\bf e}_{y}}{\omega }\sin{(\omega t)}$. The transverse mean velocity can be expanded into
\begin{equation} \label{eq:currents-AC}
\bar{v}_{x}^{\mathrm{Hall}}(t) = -\frac{\omega}{N}\sum_{n=0}^{\infty}\frac{\sin^{n}{(\omega t)}\cos (\omega t)}{n!}D_{zy}^{(n)}\left(\frac{F_{y}}{\omega}\right)^{n+1}.
\end{equation}
If the system has TRS, $D_{zy}^{(n)}$ vanishes for even integers of $n$. Hence, the period of the oscillations of the Hall response in the case of TRS is half of the period of the driving field oscillations, $T_{\mathrm{dr}}/2=\pi/\omega$. If TRS is broken, the period of the Hall response oscillations is equal to the one of the driving field. 
\begin{figure}[t]
\centering
\includegraphics[width=0.48\textwidth]{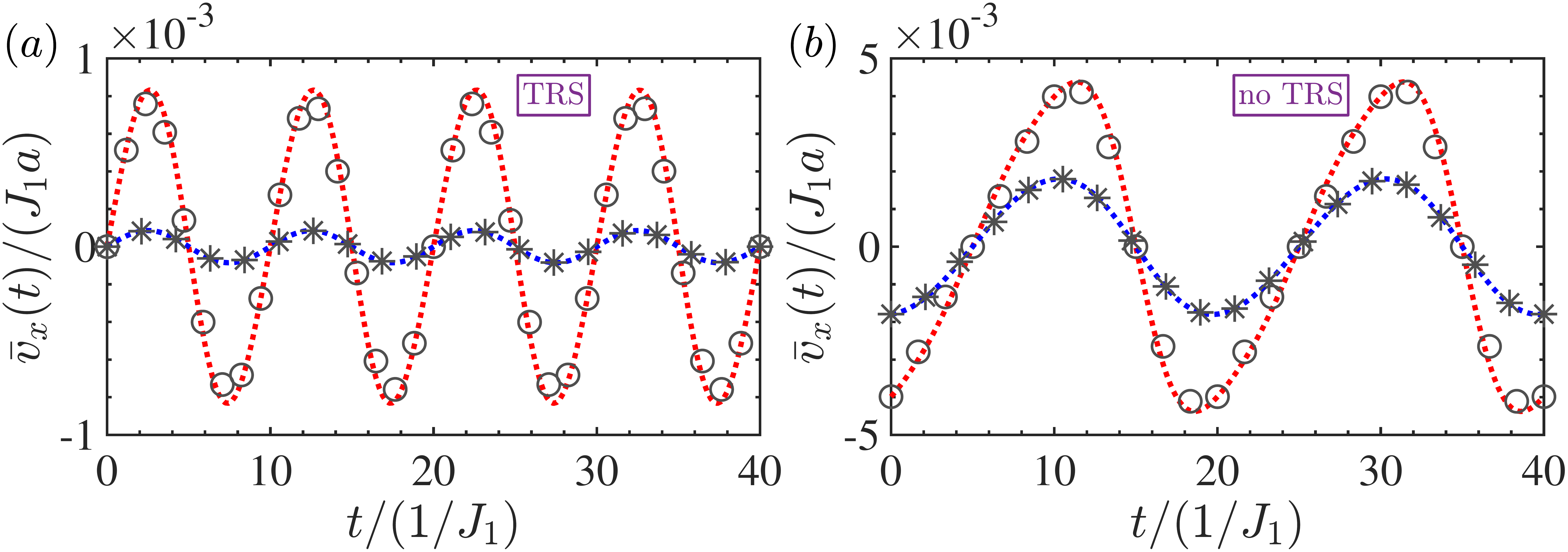} \\
\includegraphics[width=0.48\textwidth]{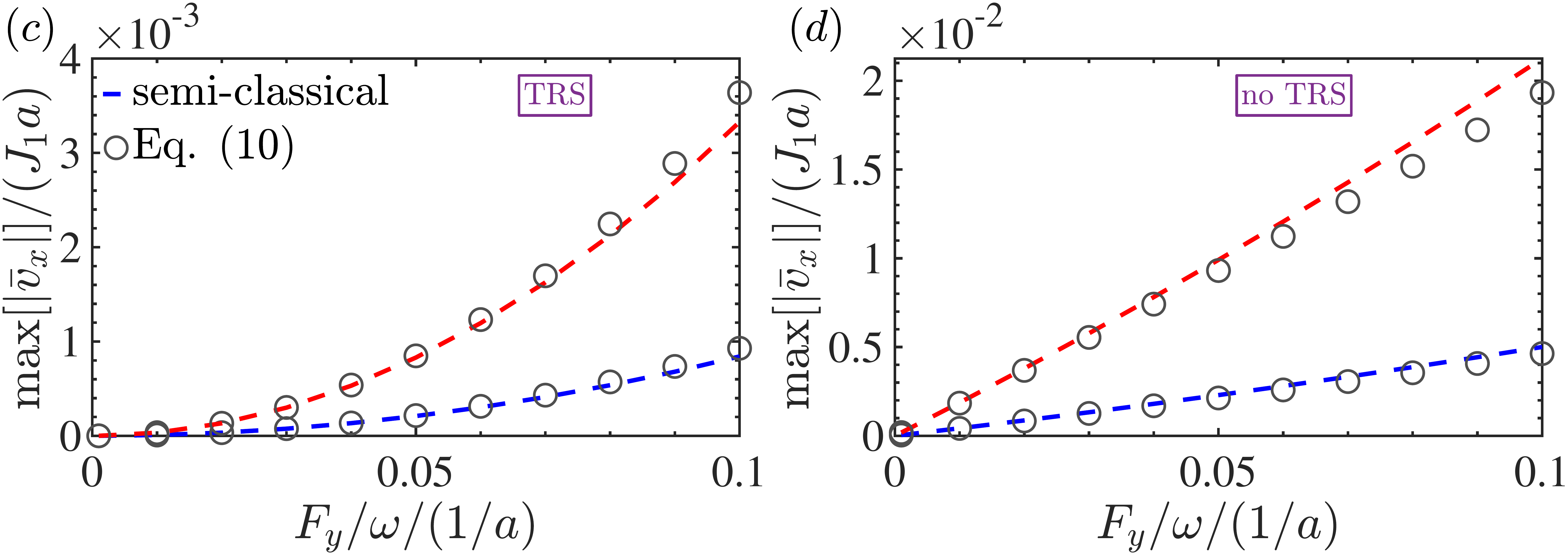}
\caption{Time-dependent mean velocity $\bar{v}_x(t)$ and its amplitude as functions of $F_y/\omega$ for oscillating force $F_y\cos{(\omega t)}$ with and without TRS. Two colors correspond to different values of $F_y/\omega$ and filling in the top and bottom panels, respectively. The lines describe the results of semiclassical calculations, while the symbols (asterisks, circles) are obtained by Eq.~\eqref{eq:vx-exp-AC}. Here, $\omega/J_1=2\pi/20$. The other parameters are the same as in Fig.~\ref{fig2}.}
\label{fig6}
\end{figure}

In the limit of small $F_{y}/\omega$ values, Eq.~\eqref{eq:currents-AC} can be expanded with retention of the second-order term as follows:
\begin{equation}
\bar{v}_{x}^{\mathrm{Hall}}(t)=\frac{1}{N}\left( \cos{(\omega
t)}D_{zy}^{(0)}F_{y}-\sin{(2\omega t)}D_{zy}^{(1)}\frac{F_{y}^{2}}{2\omega}\right).  \label{eq:vx-exp-AC}
\end{equation}
In Fig.~\ref{fig6}, we plot the mean Hall velocity $\bar{v}_{x}$ calculated from semiclassical dynamics as well as by expression~\eqref{eq:vx-exp-AC} with and without TRS. We notice that the Hall response to the oscillating longitudinal driving force in the limit of small $F_{y}/\omega$ values is almost harmonic (blue dots), and can be well described by expression~\eqref{eq:vx-exp-AC} as shown by the asterisks in Figs.~\ref{fig6}(a) and~\ref{fig6}(b). When TRS is broken, $D_{zy}^{(0)}\neq 0$, the linear part dominates in the Hall response. The maximum amplitude of oscillating transverse velocity will grow linearly with the driving force, as shown in Fig.~\ref{fig6}(d). In the systems with TRS, $D_{zy}^{(0)}=0$.
The amplitude of the Hall velocity is a parabolic function of the driving strength, as depicted in Fig.~\ref{fig6}(c). This enables us to extract the Berry curvature dipole directly from the amplitude of the transverse mean velocity.

\section{Conclusions \label{sec:summary}}
In summary, we found that the nonlinear Hall effect naturally appears in Bloch oscillations of ultracold atoms. The semiclassical dynamics reveals different behaviors with and without time-reversal symmetry due to distinct leading orders of the Hall effects. The Berry curvature dipole tensor and even the Berry curvature multipole tensor quantifying the nonlinear Hall effect can be extracted from the ac and dc driven dynamics of atoms. Current experiments with ultracold fermions could be promising to test the prediction of the nonlinear Hall effect in the Bloch oscillations.

\begin{acknowledgements}
 We acknowledge fruitful discussions with Huitao Shen and Hui Zhai. Our research was supported through the Science Foundation of Zhejiang Sci-Tech University (ZSTU) No. 21062339-Y and China Postdoctoral Science Foundation Grant No. 2020M680495, and the Beijing Outstanding Young Scientist Program held by Hui Zhai.
\end{acknowledgements}

\bibliographystyle{apsrev4-1}
\bibliography{nonlinearHall}

\end{document}